# Meissner response of superconductors: Quantum non-locality vs. quasi-local measurements in the conditions of the Aharonov-Bohm effect

Armen M. Gulian

Advanced Physics Laboratory, Chapman University, 15202 Dino Dr., Burtonsville, MD 20866

*Dedicated to Prof. Y. Aharonov on his 80th birthday*

*Abstract* – Theoretical explanation of the Meissner effect involves proportionality between current density and vector potential [1], which has many deep consequences. Amongst them, one can speculate that superconductors in a magnetic field "find an equilibrium state where the sum of kinetic and magnetic energies is minimum" and this state "corresponds to the expulsion of the magnetic field" [2]. This statement still leaves an open question: from which source is superconducting current acquiring its kinetic energy? A naïve answer, perhaps, is from the energy of the magnetic field.

However, one can consider situations (Aharonov-Bohm effect), where the classical magnetic field is absent in the space area where the current is being set up. Experiments demonstrate [3] that despite the local absence of magnetic field, current is, nevertheless, building up. From what source is it acquiring its energy then? Locally, only a vector potential is present. How does the vector potential facilitate the formation of the current?

Is the current formation a result of truly non-local quantum action, or does the local action of the vector potential have experimental consequences, which are measurable quasi-locally?

We discuss possible experiments with a hybrid normal-metal superconductor circuitry, which can clarify this puzzling situation. Experimental answers would be important for further theoretical developments.

## 1. Introduction

The question, which is discussed in this article, is whether or not it is possible to detect locally (or quasi-locally) the so-called Aharonov-Bohm potential. Experimentally, this potential, which corresponds to the local absence of the magnetic field, has been proven to act on quantum objects such as superconductors. The majority believes that its action has a non-local character. However, in physics the truth does not necessarily correspond to the opinion of the majority, but rather is determined by experiments. We propose two sets of experiments which can answer this question. Finding the right answer may have very important corollaries both for quantum physics and certain practical applications. Superconductors occupy a peculiar position among solid-state objects: they demonstrate quantum properties not only at the microscopic and mesoscopic levels, but also – and most importantly – at the macroscopic level. One encounters this feature when considering superconductors main property: charge transport without resistance. It turns out [1] (F. London and H. London, 1935) that one should adopt a relation (we drop unimportant factors in intermediate expressions):

$$j \propto -A \qquad (1)$$

to avoid working with infinitely large values of conductivity and to properly describe the Meissner effect. This immediately separates superconductors from classical objects: in classical physics observables cannot depend on *A*, since *A* is not gauge invariant. As soon as superconductors belong to the quantum world, they can supply a quantity, which will make the basic equation (1) gauge invariant. Such a quantity is the phase $\theta$ of the quantum wave function:

$$\psi = |\psi|\exp(i\theta) \qquad (2)$$

At the gauge transformation $A \to A + \nabla f$ with an arbitrary function $f(x,y,z,t)$, the phase transforms as $\theta \to \theta + f$, so that $j \propto -A + \nabla \theta$ is then gauge invariant. Does this reveal "sensitivity" of the superconductor to the vector potential? If so, then how could one measure the response of superconductors to the value of $A$? These questions are beyond pure theoretical analysis. In typical situations, when one uses superconductor-based electronics, such as SQUIDs, the responses are being expressed in terms of magnetic fluxes, $\phi$. That is easily doable using Stokes' theorem, according to which

$$\oint_C A \cdot dl = \int_S \mathrm{curl} A \cdot ds = \int_S H \cdot ds = \phi \qquad (3)$$

where $S$ is a surface bounded by the contour $C$. However, SQUIDs are closed loop systems. More generally, measurement of magnetic fields $H$ and corresponding fluxes is always possible using closed superconducting current trajectories. Then measurements are categorized as "non-local" responses. Non-locality is at the heart of quantum mechanics. The task here is not to debate that principle. The intention is to find out whether or not superconductors react locally to the presence of the vector potential.

Electromagnetic field theory operates with the local fields $E$ and $H$. It is customary to express these fields via the 4-vector potential $A_i = \{\varphi, A\}$, $i=0, 1,2,3$: $H = \nabla \times A$, $E = \nabla \varphi - \partial A/\partial t$. In classical electrodynamics the value of $A$ is defined up to the gradient of an arbitrary (gauge) function $f(x,y,z,t)$. Indeed, one can always perform a transformation $A = A' + \nabla f$, which will change neither the magnetic, nor the electric field provided the scalar potential $\varphi$ is also transformed as $\varphi = \varphi' - \partial f / \partial t$.

In quantum theory, this function $f$ couples with the phase $\theta$ of the wave function (2). As was mentioned above, this opens opportunities to measure the value of the vector potential quasi-locally using specific devices with quantum elements. We will describe the details below.

## 2. The Aharonov-Bohm potential

In the Coulomb gauge ($\mathrm{div}\, A = 0$) the static potential of the infinitely long solenoid has the form [4]:

$$A_{int}(x,y,z) = (\alpha/R^2)\{-y, x, 0\}, \quad x^2+y^2 < R^2 \text{ (internal region)}$$
$$A_{ext}(x,y,z) = \alpha\{-y/(x^2+y^2), x/(x^2+y^2), 0\}, \quad x^2+y^2 > R^2 \text{ (external region)} \qquad (4)$$

(here, the Cartesian reference frame $(x,y,z)$ is aligned with the $z$-axis coinciding with the axis of solenoid, $R$ is the radius of the solenoid, and $\alpha = H_0 R^2/2$, where $H_0$ is the magnetic field amplitude inside the solenoid directed along the z-axis). Note that the $A$-field inside the solenoid corresponds to a uniform magnetic field in the space (see, e.g., [5]), so that $\nabla \times A_{int} = H_0$. Of immediate interest is the external field described by $A_{ext}$. Direct calculation yields $\nabla \times A_{int} = 0$ at any point outside the solenoid, so there is no magnetic field $H$ associated locally with this potential $A_{ext}$. In cylindrical coordinates, the $A_{ext}$ has a very simple form:

$$A_{ext}(z,\rho,\theta) = \{A_z, A_\rho, A_\theta\} = (0, 0, \alpha/\rho). \qquad (5)$$

We will use these forms (4) and (5) in subsequent analysis.

## 3. Basic idea of the measuring device

Let us first consider a loop in the plane orthogonal to the axis of the solenoid (Fig. 1).

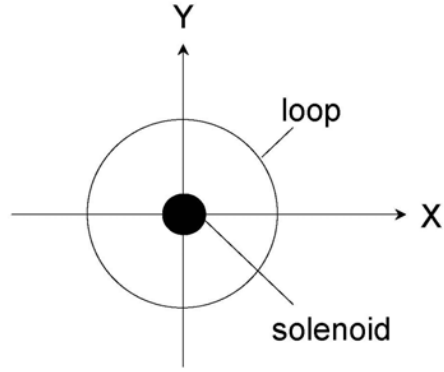

FIG. 1. Circular loop in the plane orthogonal to the axis of infinitely long solenoid. $|A_\theta(\rho)|$ is constant along the loop.

Using this figure one can describe a simple device for measuring the *A*-potential. Suppose that the loop is made of a superconducting wire. We consider a very thin wire so that the motion of the electrons is quasi-one dimensional and homogeneous throughout its cross section (*i.e.*, the diameter of the wire is much smaller than the London penetration depth λ). Then, if the loop is cooled down below the superconducting transition temperature $T_c$ in a pre-existent solenoid field, a stationary supercurrent will settle in the wire. This effect has been demonstrated experimentally.
     It is possible to detect the existence of a current in the looped wire by classical instruments via the magnetic field, which it induces in the vicinity of the wire. The whole system represents then a primitive detector of the *A*-potential. However, it may also be considered as a measuring device of a non-local influence of the magnetic field $H_0$ inside of the solenoid onto the quantum object, which is the superconducting loop: the topology of the detector allows us to integrate along the wire (the closed contour C in Fig. 1), and express the answer via the magnetic flux inside the solenoid. To exclude such an option one should not work with a closed loop.
     To do so, let us consider a finite superconducting bar, such as the one shown between points *1* and *2* in Fig. 2.

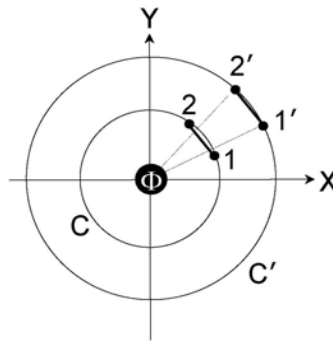

FIG. 2. A superconducting bar in the field of the solenoid. The phase difference created by the *A*-field at the ends of the bar is proportional to the angle subtended by the chord at the axis of the solenoid: $|\delta\Theta_{12}| > |\delta\Theta_{1'2'}|$.

The current density in superconducting wire is described by the usual (Ginzburg-Landau-type) equation (see, *e.g.*, [6]):

$$j = j_s = (\hbar e^*/m^*)(\nabla \theta - A) \tag{6}$$

which is a standard quantum-mechanical expression for the current density of charged particles (Cooper pairs) in the magnetic field [7] described by the wave-function $\psi$, normalized by the density $n_s$ of superconducting electrons: $|\psi|^2 = n_s$. The velocity of superfluid motion is

$$v_s = \frac{\hbar}{m^*}\left(\nabla \Theta - \frac{e^*}{\hbar c} A\right) \tag{7}$$

and the effective charge and mass are $e^* = 2e$ and $m^* = 2m$. In the 1D-motion of electrons, assuming a constant cross section for the conductor, one can deal with the current density instead of the current (*i.e.*, $|\psi|$=const, and $|A|$ and $|j|$ are constant on the contour *C*). Since *j* is obviously zero, $A = (\hbar c/e^*)\nabla \theta$, or in other words, the bar possesses a phase difference at its ends $\delta\theta = \int_1^2 A \cdot dl$, which compensates the influence of the *A*-field.

How should one measure this phase difference $\delta\theta$? A direct attempt to do so could be to move the chord with some velocity from the position *1-2* to a position *1'-2'* (as shown in Fig. 2). This will incur $\partial\delta\theta/\partial t \neq 0$, since the *A*-potential is inversely proportional to the distance from the z-axis (Eqs. 4 and 5). Oscillating the bar between these two positions one may hope to register an *AC* voltage *V* in view of the Josephson relation:

$$\partial\delta\theta/\partial t = (e/\pi\hbar)V. \tag{8}$$

This experiment, though solves the task in principle, involves mechanical oscillations, and thus seems a bit impractical. We consider next two more ideas, which look feasible experimentally. However, the introduced example is very important, since it allows recognizing properly the role of the superconducting element. In the Coulomb gauge, in the co-moving reference frame in which the oscillating bar is in rest, we have $\partial A/\partial t \neq 0$. Be this value a gauge-invariant one, we would have electric field $E = -\partial A/\partial t$ acting on the bar, and detectable by a current it would have caused in a normal bar. However, by the choice of an appropriate gauge function f(x,y,z,t) the *A*-field may be zeroed in the co-moving reference frame, so that $A' \equiv 0$, and $\partial A'/\partial t = E = 0$. The situation with a superconducting bar is quite different. In the gauge $A \equiv 0$, it is not the *A*-function which creates $\partial\delta\theta/\partial t \neq 0$, but the oscillating gauge function f(t) which contributes to the phase of the quantum wave function (2).

## 4. First experimental approach: oscillating fields

Let us again consider a superconducting bar (*1-2*) on a circumference as shown in Fig. 2. Let this bar be a part of a normal circuit containing classical measurement devices. These external devices put current into the circuit, and are able to measure the energy required for the motion of charged carriers. How much energy is required to put a current *j* in a superconducting bar provided current passage is allowed at the ends, say, the bar is a part of a closed loop around the solenoid? The required energy is related to the so-called kinetic inductance, which is characteristic of superconductors at any frequency. For the energy density $\varepsilon$ one has [8]:

$$\varepsilon = \Lambda j^2/2 = n_s m v_s^2/2 \quad , \tag{9}$$

where $\Lambda$ is the kinetic inductance. If the superconducting bar has a thickness $\lambda_0$, and a height $w$, its kinetic inductance energy is $\delta E = \varepsilon \lambda_0 w \int_1^2 \mathbf{A} \cdot d\mathbf{l}/|\mathbf{A}|$, where $\varepsilon$ is related to the *A*-potential via Eq.(9) and $v_s = -(e^*/m^*c)\mathbf{A}$, which follows from Eq.(6) in the case of unrestricted motion of the supercurrent ($\nabla \theta = 0$). Our suggestion is that the external current source mentioned above mimics the unrestricted motion by removing charges from the ends of the bar. If there is *no flux* in the solenoid, the current source provides the energy (7) when switched on, and sets up a superfluid motion in the bar.

Suppose now that there *is a flux* $\phi$ in the solenoid (and a corresponding *A*-field, Fig. 1). Because of the action of the *A*-field, at a fixed value of current *j* in our measuring device, the contribution of the external source should be less by an amount of $\delta E$, if the direction of the current it is initiating matches the direction of flow being initiated by the *A*-field. Reciprocally, the source should provide $\delta E$ more energy if the directions are opposite. By the same logic, in the case of symmetric AC excitation of the device, the presence of the *A*-field will cause an asymmetric response in the circuit (Fig. 3).

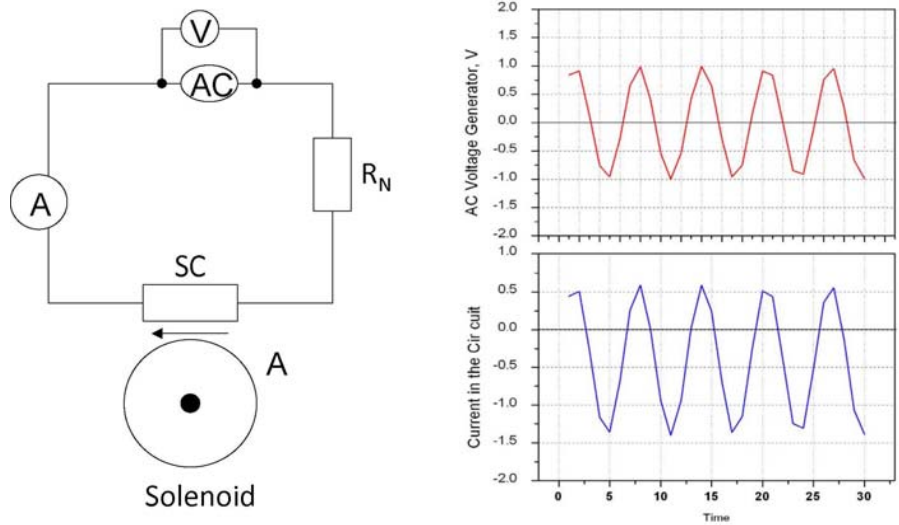

FIG. 3. Possible scheme for a detection of the *A*-potential. If the source is a voltage source with a sinusoidal AC output, then during consecutive half-periods different electromotive forces are acting in the circuit. Then the current is not symmetric (we plotted a case when the *A*-potential is opposing positive values of the current). All components except the superconducting bar are classical (non-quantum) elements. We emphasize that the presence of a quantum element in the overall scheme (the superconducting bar in our case) is mandatory.

## 5. Second experimental approach: pulsed fields

Our second approach is a pulse-type process. A strip of a superconductor with a cross-section *S* and length *L* is placed in the AB-field (Fig. 2), so that the *A*-vector is either parallel or anti-parallel to the length *L* (Fig. 4).

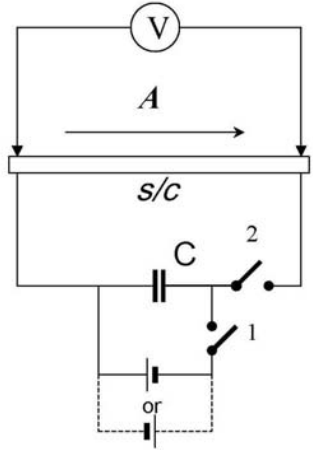

FIG. 4. Schematics of the pulse detection of the AB potential.

For a given orientation and value of the *A*-potential due to the solenoid, one charges the capacitor *C* to a level *Q* (with a chosen voltage *V* and polarity), then opens switch 1 to disconnect the voltage source, and closes switch 2, thus putting current trough the superconducting strip. If the current is less than the critical value, then voltmeter will show no voltage. One can repeat the operations, and increase the voltage, thus increasing the energy $E_C=Q^2/2C=CU^2/2$, stored in the capacitor. When the energy reaches the amount required to destroy the superconducting state in the strip, one stops, and repeats the same procedure with the reversed polarity of the battery. Asymmetry between the threshold values will reveal the role of the vector potential *A* thus detecting it quasi-locally.

## 6. Discussion

To understand how close these methods are to practical realization, let us make some estimates.

In the case of oscillating field, performing integration, one obtains $\delta E = n_s m v_s^2 \lambda_0 w\,[f(2) - f(1)]\,r/2$, where $f(x,y) = -\alpha\,\arctan(x/y)$. Also, $v_s^2 = (\alpha^2/r^2)(e^2/m^2c^2) = (e^2\Phi^2/4\pi^2 r^2 m^2 c^2)$. Combining these results, one gets: $\delta E = (e^2 H_0^2 R^4 n_s \lambda_0 w \vartheta / 8r\,mc^2)$, where $\vartheta = f(2) - f(1)$. Let us estimate the value of this energy. We choose $\vartheta \sim 1$, $R \sim 0.1$ cm, $r \sim 3R$, $\lambda_0 \sim 10^{-5}$ cm, $w \sim 10^{-5}$ cm, $n_s \sim 10^{23}$ cm$^{-3}$, and $H_0 \sim 10$ Oe. Then $\delta E \sim 1$ nJ, which is a quite detectable value. For comparison, the energy of the so-called RSFQ pulses used in superconducting electronics is ~0.2 aJ, *i.e.*, $\delta E$ is equivalent to five billion RSFQ pulses. In addition, averaging over many cycles is feasible, which makes $\delta E$ easier to detect.

In case of pulsed fields, suppose the bar has $L=1000\,\mu m$ and $S=0.01\,\mu m^2$ as in the previous case. Then the volume is $10^{-11}$ cm$^3$ and the number of Cooper pairs is about $10^{12}$. Each Cooper pair has ~1 meV energy, which means that the upper limit to the energy stored in the capacitor should be about $10^9$ eV, or about *100 pJ*. One should expect delectability of the effect by the use of simple equipment shown in Fig. 4. In principle, the energy can be delivered using a more sophisticated device, such as an energy pulsing electronics circuit.

## Conclusions

We have described conceptually simple experiments, which can be used to detect the local influence of the Aharonov-Bohm vector potential on a quantum object – superconductor. The outcome of these experiments can demonstrate whether or not quantum systems can sense the vector potential locally. A positive answer would affect basics of quantum electrodynamics, while a negative answer would clarify more the non-local action in quantum physics. Both cases would serve the perplexed [9].

*Acknowledgement* – I am very grateful to Yakir Aharonov, Jeff Tollaksen, Alan Kadin, Lou Sica, Joe Foreman, Michael Steiner, and Mamikon Gulian for numerous discussions.